\documentstyle[preprint,aps,floats,epsfig]{revtex}
\begin{document}
\title{Non magnetic molecular Jahn-Teller Mott insulators}
\author{Michele Fabrizio$^\dagger$ and Erio Tosatti$^{\dagger,\star }$}
\address{$^\dagger$ Istituto Nazionale di Fisica della Materia I.N.F.M.,}
\address{and International School for Advanced Studies (SISSA),
via Beirut 2-4, I-34014 Trieste, Italy,}
\address{$^\star$ International Center for Theoretical Physics 
(ICTP), Trieste, Italy}
\date{July 2, 1996}
\maketitle
\begin{abstract}
Narrow-band conductors may turn 
insulating and magnetic as a consequence of strong electron-electron 
correlation. In molecular conductors, the concomitance 
of a strong Jahn-Teller coupling may give rise to
the alternative possibility of a non-magnetic insulator, 
with or without a static cooperative Jahn-Teller distortion.
In the latter case the insulator has Mott-like properties,
with an interesting interplay between electron-electron repulsion and the
Jahn-Teller effect, which is dynamical. 
We study this kind of non-magnetic insulator
in a very simple $E\otimes e$ Jahn-Teller model and we discuss
its general properties in a more general context, also in connection
with the insulating state of K$_{4}$C$_{60}$ and Rb$_{4}$C$_{60}$. 
\end{abstract}


\section{Introduction}
The existence of a model of conducting electrons which turns into 
a non trivial paramagnetic Mott insulator at zero 
temperature is a long standing question.
The problem can be formulated in the following way.
Let us consider a model which
is supposed to be a metal if some kind of correlation (among the electrons
or between the electrons and the ions) is neglected. Can these correlations
turn the model into a Mott insulator, as opposed to a 
band insulator?
(For a band insulator we mean a system which fits all the standard
one-electron criteria of an insulator, i.e. an even number of valence
electrons per unit cell, an integer number of filled band, and a band gap.) 
In fact, in spite of many efforts, many of the 
models which have been proposed to this end become insulating
by enlarging the unit cell and consequently 
reducing the Brillouin zone so to fit those criteria.   
Typical examples are the Hubbard model in $D>1$ or
the one-dimensional 
Su-Schrieffer-Hegger (SSH) models
at half filling. In these cases the density corresponds to one valence 
electron per unit cell, so the models should be in principle metallic. 
However,
the electron-electron interaction in the Hubbard model and the 
electron-phonon coupling in the SSH model force on the system a staggered 
order parameter which doubles the unit cell, so that
the systems acquire the pre-requisite for being insulators, and in fact
they become insulating.

In this paper we are going to describe an alternative kind of
Mott insulator. Let us consider a lattice made of molecules 
which have unfilled degenerate valence orbitals occupied by an 
\underline{even} number
of electrons.  Let us briefly discuss how the lattice of such
molecules might be insulating. A first trivial possibility is
that the crystal field and the electron hopping split the 
degeneracy so to give rise to a
band insulator once filled by the available electrons.  
We are not going to consider this case. That is we will assume that, even
after the crystal field and the hopping are duly taken into account, the 
resulting band structure still implies a metallic state. 
In such a situation, considering electron-electron repulsion, a Mott
insulating behavior may still occur
because the energy cost to change the valence of a molecule by
$\pm 1$ electron overcomes the gain in energy due to the 
electron hopping. Specifically, if the isolated molecule has
$2n$ valence electrons and we define an effective Hubbard $U_{eff}$ as
\begin{equation}
2 U_{eff} = E(2n+1)+E(2n-1)-2E(2n),
\label{U-eff}
\end{equation}
where $E(m)$ is the molecular ground state energy with $m$ valence
electrons, a Mott state can be expected for $U_{eff}\gg W$, where
$W$ is the electron bandwidth.

However, in this case we still have two possible routes to an insulating state.
If the molecular ground state has a non-zero spin due to a prevailing
Hund's rule coupling, the insulator
will likely be magnetic, similarly to the previously discussed
Hubbard model at half filling. In particular, the neighboring molecules
will couple either ferro- or antiferromagnetically depending upon the 
detailed properties of the system. 
  
In the alternative case where the molecular ground state
is a singlet, we expect instead a non-magnetic Mott insulator.
A possible mechanism for a Hund's rule violation, leading to a singlet in
a molecule with partially filled orbitals, is of course provided by the
Jahn-Teller (JT) effect. Let us therefore assume that the
coupling to some vibrational mode is able to split the degeneracy
of the molecular orbitals favoring, in the absence of Coulomb repulsion,
a non degenerate singlet state. In the following we will consider a coupling 
to a local vibron, driving the single molecule
to a Jahn-Teller distortion. Then, if the JT 
energy splitting $\Delta E_{JT}$
between the non degenerate singlet and the high-spin states 
overcomes the Hund's rule (Coulomb) splitting $\Delta E_{Hund}$, 
then the molecular ground state 
is the non degenerate singlet and the insulator 
is non magnetic\cite{Khomskii}. 
[Notice that the JT effect will in general also contribute to 
an increase of $U_{eff}$ in Eq.(\ref{U-eff}).] 

Now, there are still two
possibilities for both the magnetic and non magnetic insulators.
If the intra-molecular coordinates can be treated as classical,
we can fix them at each site to some values and then
solve the electron problem in the external potential provided by the 
static molecular distortions. 
The ground state lattice configuration is found by minimizing the 
total energy given by the vibrational energy plus the electron energy (in
the presence of the molecular deformation), and in general consists
of a periodic arrangement of statically distorted molecules (cooperative
JT effect). Note that a commensurate distortion 
might turn the metallic state into an insulator even if
$U_{eff}\sim W$ as a consequence of a nested Fermi surface. 

A second and newer possibility arises if the quantum
fluctuations of the intramolecular coordinates  
cannot be neglected. This is the case, for example, with 
a large spacing between the lowest rotational levels in the Jahn-Teller
trough potential (or the tunnel splitting in a warped one) 
compared with the intermolecular coupling.
In this case a static collective molecular distortion may become 
disadvantageous and the rotational (or cubic, or bct, etc...) symmetry of each 
molecule dynamically restored (dynamic JT effect). This would be a 
quite interesting situation
in which the Jahn-Teller effect plays a crucial role in making 
the insulating state non-magnetic without revealing itself by
a cooperative distortion. We will define this insulator as a
Jahn-Teller-Mott (JTM) insulator.
 
Hence, a lattice of molecules with this kind of properties
can in principle displays a variety of behavior depending upon
the relative strengths of the parameters which come into play.
If $U_{eff}\gg W$, we expect an insulating behavior either
magnetic, if $\Delta E_{Hund} \gg \Delta E_{JT}$, or non magnetic,
if $\Delta E_{JT}\gg \Delta E_{Hund}$. On the contrary, if $W \gg U_{eff}$,
a metallic state should be favored. A cooperative
JT effect may or may not be present depending on the
value of the lowest rotational frequencies compared to the intermolecular
coupling. In the particular case of $\Delta E_{JT}\gg \Delta E_{Hund}$ we 
expect that, as the bandwidth increases, a transition should take place from
a non magnetic insulator to a metal.

A source of inspiration for this work has been the fullerides 
$K_4 C_{60}$ and $Rb_4 C_{60}$, which, as we discuss in section IV,
have many features in common with the general model we have previously 
discussed, although in these materials it is still
unclear from the experimental and theoretical point of views which of the
above introduced scenarios is effectively realized.
However, our main purpose here is to introduce and discuss a new mechanism for 
the occurrence of a non-magnetic Mott insulator, which may, 
in way of principle, be realized in a different system, even though not yet 
found or studied. Therefore, since the situation in fullerenes 
is still open,
we have postponed the discussion of the possible realization of this 
insulating state in those materials to section IV. 
We instead use the following two sections to study a prototype simple
model which does show all the interesting features we have previously
foreshadowed, depending on the values of the different 
parameters in the Hamiltonian. Specifically, this model consists of a lattice
of ``molecules'', each with two degenerate orbitals coupled to
a doubly degenerate vibron ($E\otimes e$ Jahn-Teller 
problem\cite{Englman}). Different molecules are then coupled via a single 
particle electron hopping term, and, in addition, an intra-molecule Coulomb
interaction is taken into account. 

\section{Single molecule}

We start by describing our modeling of the isolated molecule.
We take the following molecular Hamiltonian ($\hbar=1$):
\begin{equation}
\hat{H}= \frac{\omega_0}{2}\left(\vec{p}\,^2 + \vec{r}\,^2 \right) +
\omega_0 g \vec{r}\cdot\vec{\tau} + \hat{H}_{int},
\label{Ham}
\end{equation}
where $\vec{r}$ is a two dimensional vibrational coordinate,
$\vec{p}$ its conjugate momentum, and $\vec{\tau}$ are pseudospin-1/2 
matrices acting in the space of the two degenerate electronic orbitals,
i.e.
\begin{eqnarray}
\tau^+ &=& \sum_\sigma c^\dagger_{1\sigma} c^{\phantom{\dagger}}_{2\sigma},\\
\tau_z &=& \frac{1}{2}\sum_\sigma \left(c^\dagger_{1\sigma}  
c^{\phantom{\dagger}}_{1\sigma}
 -  c^\dagger_{2\sigma}  c^{\phantom{\dagger}}_{2\sigma}\right).
\end{eqnarray}
The interaction part $\hat{H}_{int}$ of the Hamiltonian can be
written in the form
\begin{equation}
\hat{H}_{int} = U n^2 + V \left( \tau_x^2 + \tau_y^2 \right),
\label{Hint}
\end{equation}
which takes into account the planar symmetry of the model.
Here $n=\sum_\sigma c^\dagger_{1\sigma} c^{\phantom{\dagger}}_{1\sigma}
+ c^\dagger_{2\sigma} c^{\phantom{\dagger}}_{2\sigma}$
is the total occupation number, and $U$, $V$ are positive interaction
parameters. In particular $U$ will limit fluctuations of total charge of
the molecule, while $V$ controls the internal spin and orbital state.

By rewriting in polar coordinates $\vec{r}=(r\cos\phi,r\sin\phi)$ the coupling
between the vibron mode and the electrons, we get
\begin{equation}
\vec{r}\cdot\vec{\tau} = \frac{1}{2} r {\rm e}^{i\phi} \tau^- + H.c.\,.
\end{equation}
The angular coordinate dependence can be partly eliminated via the 
(non-abelian) gauge transformation:
\[
\hat{H} \mapsto \hat{U} \hat{H} \hat{U}^\dagger,
\]
where
\begin{equation}
\hat{U} = {\rm e}^{i\phi\tau_z}.
\label{U}
\end{equation}
Under this transformation
\begin{eqnarray}
c^{\phantom{\dagger}}_{1\sigma}&\mapsto& 
c^{\phantom{\dagger}}_{1\sigma} {\rm e}^{-i\phi/2},\nonumber \\
c^{\phantom{\dagger}}_{2\sigma}&\mapsto& 
c^{\phantom{\dagger}}_{2\sigma} {\rm e}^{i\phi/2},
\label{gauge}
\end{eqnarray}
so that the Hamiltonian (\ref{Ham}) simplifies into:
\begin{equation}
\hat{H}= -\frac{\omega_0}{2r}\frac{\partial}{\partial r}
\left( r \frac{\partial}{\partial r}\right)
-\frac{\omega_0}{2r^2} \left( \frac{\partial}{\partial \phi} + i\tau_z
\right)^2 + \frac{\omega_0}{2}r^2 + \omega_0 g r \tau_x + \hat{H}_{int}.
\label{Hambis}
\end{equation}
The angular part of (\ref{Hambis}) is simply diagonalized by
eigenfunctions $\exp{(i\lambda \phi)}$. Moreover, since the 
total wavefunction must not change upon varying $\phi$ by $2\pi$, as a 
consequence of the
gauge transformation (\ref{U}) $\lambda$ has to be integer for
even electron number and half odd integer for odd electron number.
Due to the symmetry $\lambda\to -\lambda$ and to the spin degeneracy, 
each level for odd electron number is at least four times degenerate
(in this case the dynamic JT effect does not split the orbital 
degeneracy).
On the contrary, for two electrons (the states with zero or with four
electrons are trivial) the degeneracy is lifted and the ground state is non 
degenerate. Let us now assume a strong JT coupling, $g\gg 1$,
and calculate the ground state and lowest energy spectrum
with an accuracy of order $g^{-2}$.
In this limit, we can adopt a Born-Oppenheimer (BO) approximation scheme
for the transformed Hamiltonian (\ref{Hambis}). In fact,
the ``electronic'' states (notice that the transformed electrons are indeed
a mixing of purely electronic and vibrational degrees of freedom), which are
not orbital-singlets, will be split by the electron-vibron coupling
of terms of order $\omega_0 g \langle r \rangle$, where 
$\langle r \rangle$ is the average value of the vibron displacement in the
BO approximation, which is different for different electronic configurations.
If $g\gg 1$, the energy differences between different electronic
configurations will be much larger than the lowest excitation energies
within the Jahn-Teller trough potential, thus justifying the
BO approximation.

We therefore start by calculating the electronic energies and eigenstates
at fixed $r$. These are listed in Table I for different occupation
numbers $n=1,2,3$, where each eigenstate is
identified by a set of quantum numbers. In particular, 
when the total orbital momentum is not zero, the appropriate
orbital quantization axis is $x$\cite{notax}. It is therefore
useful to introduce the rotated orbitals 
which diagonalize $\tau_{x}$, defined by 
\begin{eqnarray}
d^{\phantom{\dagger}}_{1\sigma} &=& 
\frac{1}{\sqrt{2}}\left( c^{\phantom{\dagger}}_{1\sigma} +
c^{\phantom{\dagger}}_{2\sigma}\right), \label{duno}\\
d^{\phantom{\dagger}}_{2\sigma} &=& 
\frac{1}{\sqrt{2}}\left( c^{\phantom{\dagger}}_{1\sigma} -
c^{\phantom{\dagger}}_{2\sigma}\right). \label{ddue}
\end{eqnarray}

The next step is to diagonalize the Hamiltonian for the vibrational coordinates
using the electronic eigenvalues $E(r)$'s of Table I as effective potentials.
We are going to discuss the final result only for the lowest energy states.

For two electrons, the molecule has two sets of low energy states. The first 
one is spin-singlets which take more advantage from the 
JT effect (see Table I).
Notice, however, that the vibrational potential energy of this state shows a 
minimum at finite $r$ only if $2\omega_0 g^2 > V$, otherwise the minimum stays
at $r=0$. 
This describes the competition on the single molecule between 
the Hund's rule, which is related to
$V$ of Eq.(\ref{Hint}), against a JT effect. If the latter 
is strong enough, so that $2\omega_0 g^2 > V$, then a distorted potential 
energy minimum develops at
\[
r_* = g \sqrt{ 1 - \left(\frac{V}{2\omega_0g^2}\right)^2},
\]
and the low energy spectrum of singlets is
\begin{eqnarray}
^1E^{(2)}_{l,\lambda} &\simeq& \left(l +\frac{1}{2}\right)\omega_* 
+ \frac{\omega_0}{2}\left(\frac{\lambda}{r_*}\right)^2 
- \frac{\omega_0}{2}g^2 - \frac{V^2}{8\omega_0 g^2} \nonumber \\ 
&+& 4U + \frac{3}{2}V,  \label{eigen-jt}
\end{eqnarray}
where $l=0,1,2,\cdots$ is the quantum number labeling the 
small oscillations in the
JT potential well, whose frequency $\omega_*$ is 
\begin{equation}
\omega_* = \omega_0 \sqrt{ 1 - \frac{V^2}{4\omega_0^2 g^4} }, 
\label{omegastar}
\end{equation}
and $\lambda=0,1,2,\cdots$ is the angular momentum
eigenvalue. In Eq.(\ref{eigen-jt}) we have omitted constant terms
of order $\geq \omega_0/g^2$ as well as $l$- and $\lambda$-dependent 
terms of higher order. 

The second set of states for the two electron molecule consists of 
spin-triplets. Since these are orbital-singlets,
the JT coupling is ineffective and the 
low energy spectrum is simply
\begin{equation}
^3E^{(2)}_{n_1,n_2} =  \left(n_1 + n_2 + 1\right)\omega_0 + 4U.
\label{eigen-T}
\end{equation} 
If the JT energy gain overcomes
the Hund's rule gain, that is if
\begin{equation}
\frac{\omega_0}{2}g^2 + \frac{V^2}{8\omega_0 g^2} > \frac{3}{2}V,
\label{condition}
\end{equation}
then the ground state is a spin-singlet belonging to the set (\ref{eigen-jt}),
and the lowest singlet excitations 
in the $g\gg 1$ limit are obtained by increasing the angular momentum
$\lambda$.  

When the molecule is occupied by one or three electrons, on the other hand,
the low energy spectrum is simply
\begin{eqnarray}
^2E^{(1,3)}_{l,\lambda} &\simeq& 
\left(s+\frac{1}{2}\right)\omega_0
+ 2 \omega_0\left(\frac{\lambda}{g}\right)^2 - \frac{\omega_0}{8}g^2
\nonumber \\
&+& U n^2 + \frac{V}{2},  \label{eigenunotre} 
\end{eqnarray}
where $n=1$ or $n=3$, respectively, and $\lambda$ is half an odd
integer.

When condition (\ref{condition}) is fulfilled, i.e. when the
$n=2$ molecular ground state is a spin-singlet,
the effective Hubbard $U_{eff}$ is given by
\begin{equation}
2 U_{eff} = {^2E^{(1)}_{1/2,0}} + {^2E^{(3)}_{1/2,0}} 
- 2\, {^1E^{(2)}_{0,0}} \simeq \frac{3}{4}\omega_0 g^2 + 2U 
+ \frac{V^2}{4\omega_0 g^2} -V.
\label{Ueff}
\end{equation} 
We are interested in the case where the above $U_{eff}$ is much bigger than 
the electron bandwidth $W$.

\section{A lattice of molecules with two electrons per site}

Let us now consider a lattice of molecules, each of them described by the
molecular Hamiltonian (\ref{Ham}), with an average electron density 
of two electrons per site (half filling). As previously discussed,
we assume that for $n=2$ the molecular ground state is a singlet due to
the Jahn-Teller effect prevailing the Hund's rule.
If the molecules were uncoupled, 
the system would be a trivial insulator made of independent
dynamic Jahn-Teller molecules each doubly occupied and non degenerate. An 
inter-molecular coupling is introduced via a single particle hopping of the 
general form
\begin{equation}
\hat{T} = -\sum_{ij} \sum_\sigma \sum_{a,b=1}^2 t^{ab}_{ij}
c^\dagger_{ia\sigma} c^{\phantom{\dagger}}_{jb\sigma} .
\label{hopping}
\end{equation}
We will assume that the hopping energy is enough smaller than
the effective Hubbard $U_{eff}$ of Eq.(\ref{Ueff}),
so that the addition of (\ref{hopping}) into the Hamiltonian
does not automatically turn the system into a metal.
Nevertheless the interplay between the on site energy and the hopping can 
in principle lead to different situations, as anticipated in the Introduction.

\subsection{Effective model in the strong Jahn-Teller coupling limit} 

In this Section we are going to derive an effective model valid in the
limit $U_{eff}\gg W$ and $g\gg 1$.

In the absence of the hopping term (\ref{hopping}), each molecule 
would be in its ground state with quantum numbers 
$(l,\lambda)=(0,0)$ and total spin $S=0$
[see Eq.(\ref{eigen-jt})]. 
The hopping mixes the molecular ground state with the excited states. 
We will concentrate only to singlet excitations which do not change 
the values of $l$, i.e. we will 
keep only the excitations to larger $\lambda$'s. To be consistent, we must
consider only excitations up to $\lambda\sim g$. 
Within these approximations, the second order perturbation theory
in (\ref{hopping}) provides, as we are going to show, an effective 
intermolecular contribution $\hat{J}$ 
to the Hamiltonian which only acts on the subspace 
with molecular quantum numbers $l=0$ and $S=0$
but different $\lambda$'s.
The matrix elements of this term are formally given by
\begin{equation}
\langle a'| \hat{J}|a\rangle = \sum_b \langle a' |
\hat{T} | b \rangle \langle b |
\hat{T} |a \rangle\frac{1}{E_{a}
-E_{b}},
\label{eff-T}
\end{equation}
where in the states $|a\rangle$ and $|a'\rangle$ all the molecules
are in eigenstates belonging to the set of Eq.(\ref{eigen-jt}) 
with $l=0$. 
After the gauge transformation (\ref{U}) is performed, (\ref{hopping})
is transformed onto:
\begin{eqnarray*}
\hat{T}  &=&  -\sum_{ij} \sum_\sigma t^{11}_{ij} {\rm e}^{i(\phi_i-\phi_j)/2}
c^\dagger_{1i\sigma} 
c^{\phantom{\dagger}}_{1j\sigma} +   
 t^{22}_{ij} {\rm e}^{-i(\phi_i-\phi_j)/2}
c^\dagger_{2i\sigma} 
c^{\phantom{\dagger}}_{2j\sigma} \\
&+& t^{12}_{ij} {\rm e}^{i(\phi_i+\phi_j)/2}
c^\dagger_{1i\sigma} 
c^{\phantom{\dagger}}_{2j\sigma} +
t^{21}_{ij} {\rm e}^{-i(\phi_i+\phi_j)/2}
c^\dagger_{2i\sigma} 
c^{\phantom{\dagger}}_{1j\sigma} \,, 
\end{eqnarray*}
which, in terms of the orbitals $d_1$ and $d_2$ of 
Eqs.(\ref{duno})-(\ref{ddue}), becomes
\begin{equation}
\hat{T} = - \sum_{ij}\sum_\sigma\sum_{a,b=1}^2
T^{ab}_{ij}d^\dagger_{a i\sigma}d^{\phantom{\dagger}}_{j b\sigma}.
\label{hopp-d}
\end{equation}
The hopping matrix elements in the above equation are
\begin{eqnarray}
2 T^{ab}_{ij}&=&  \left(
  t^{11}_{ij} {\rm e}^{i(\phi_i-\phi_j)/2}
+ t^{22}_{ij} {\rm e}^{-i(\phi_i-\phi_j)/2}\right) \delta_{ab}
+ \left(
  t^{11}_{ij} {\rm e}^{i(\phi_i-\phi_j)/2}
 -t^{22}_{ij} {\rm e}^{-i(\phi_i-\phi_j)/2}\right) \sigma^x_{ab}
\nonumber \\
&+& \left( t^{12}_{ij} {\rm e}^{i(\phi_i+\phi_j)/2}
+ t^{21}_{ij} {\rm e}^{-i(\phi_i+\phi_j)/2}\right) \sigma^z_{ab}
-i \left( t^{12}_{ij} {\rm e}^{i(\phi_i+\phi_j)/2}
- t^{21}_{ij} {\rm e}^{-i(\phi_i+\phi_j)/2}\right) \sigma^y_{ab}.
\label{TT}
\end{eqnarray}
Both our initial and final molecular states, $|a\rangle$ and $|a'\rangle$
in Eq.(\ref{eff-T})  have the 
orbitals $d_2$ of Eq.(\ref{ddue}) doubly occupied at each site. 
Therefore the only processes 
allowed at second order by the hopping correspond to move an electron 
from an orbital $d_2$ at site $i$ into a $d_1$ orbital at site $j$ 
(which is the only one available) or 
vice versa, and then move it back. 
The sum over the intermediate states $|b\rangle$ 
in Eq.(\ref{eff-T}) only runs over molecular configurations in which 
two molecules are in a state with odd number of particles
and in general with $l\geq 0$, $\lambda\geq 0$. 
In the $U_{eff}\gg (W,\omega_0)$ limit, the energy denominator in 
Eq.(\ref{eff-T}) can be taken as a
constant $\simeq - 2U_{eff}$
for any $|b\rangle$. Consequently the sum over $|b\rangle$ 
becomes a completeness, which implies for instance that the hopping is
not renormalized by any Ham reduction factor. 
Under all the previous assumptions,
the explicit expression of (\ref{eff-T}) reads:
\begin{equation}
\hat{J}= -\frac{1}{2U_{eff}}\sum_{ij} \left(T^{12}_{ij}\right)^* T^{12}_{ij}
+  T^{21}_{ij} \left(T^{21}_{ij}\right)^*,
\end{equation}
which, through Eq.(\ref{TT}) and apart from constant terms, is equal to 
\begin{equation}
\hat{J} = \frac{1}{8U_{eff}} \sum_{ij} 
\left( t^{11*}_{ij} t^{22}_{ij} {\rm e}^{-i(\phi_i -\phi_j)}
+ c.c.\right) + 
\left(t^{12*}_{ij} t^{21}_{ij} {\rm e}^{-i(\phi_i+\phi_j)} + c.c.\right).
\label{Exc}
\end{equation}
Notice that an analogous intermolecular phase coupling would also arise from
multipolar forces among the molecules. We will
not take these forces explicitly into account, even though they are as well
important, since their
inclusion does not modifies the following qualitative discussion.
         
In addition to the intermolecular interaction (\ref{Exc}), we must add the
intramolecular orbital energy $\hat{H}_0$, which in the same $g\gg 1$ limit is
\begin{equation}
\hat{H}_0= -\frac{\omega_0}{2 r_*^2} \sum_i 
\frac{\partial^2}{\partial \phi_i^2}.
\label{onsite}
\end{equation}
The sum of (\ref{Exc}) and (\ref{onsite}) thus represents the effective
Hamiltonian acting on the \underline{phases} of the molecular vibrons at 
leading
order in $1/U_{eff}$ and $1/g^2$. On one hand, the intersite coupling
(\ref{Exc}) tends to fix statically the phase of each molecule to get 
advantage from 
the electron hopping (cooperative Jahn-Teller effect). 
On the other hand, the on site rotational energy
favors a state in which the angular momentum is fixed and
consequently the phase is indeterminate (dynamical Jahn-Teller effect). 

From the above discussion we expect that,
as the intermolecular interaction increases
with respect to the lowest rotational frequency, 
the model should undergo a transition
from a state where the vibron phase is disordered to a state where it orders.
In both cases the system is insulating.
 
The transition to a metallic phase can not be described in this 
scheme, which is valid only in the limit of bandwidth much smaller than
$U_{eff}$.

\subsection{Excitations in the insulating  phase}

A possible way of discriminating among the two kinds of non magnetic 
insulator states, one which is accompanied by a cooperative JT effect 
and another one which is not, is through the study of the low energy
excitations. 

Let us start with the dynamical case, i.e. let us assume 
that the rotational frequency overcomes the intermolecular coupling.

The first kind of excitation we consider is a neutral spin-1 exciton.
A single molecule is in a triplet state with, for instance, $S_z=1$ 
and it has both orbitals singly occupied with a spin up electron.
No JT effect occurs in this case.
All the other molecules are in the
singlet molecular ground state with $l=\lambda=0$
[see Eq.(\ref{eigen-jt})]. This state is obviously degenerate since any
of the molecules can be in the triplet state. This degeneracy
can be lifted by the single particle hopping via a second order
process (see Fig.1). 
As a consequence the triplet excitation moves
coherently with an exciton hopping Hamiltonian given by
\begin{equation}
\frac{\Gamma}{2U_{eff}} 
 \sum_{ij} \langle T^{11}_{ij}T^{22}_{ji}\rangle  | i\rangle\langle j |
\label{hopp-S}
\end{equation}
where the average $\langle \cdots \rangle$ is over molecular states
with $\lambda=0$, and gives
\[
\langle T^{11}_{ij}T^{22}_{ji}\rangle = \frac{1}{4}
\left( |t^{11}_{ij}|^2 + |t^{22}_{ij}|^2 - |t^{12}_{ij}|^2 
-  |t^{21}_{ij}|^2\right).
\]  
The constant $\Gamma$ in Eq.(\ref{hopp-S}) is a Ham's factor reducing
the bandwidth of the spin-1 excitations and comes from the overlap
of the vibronic wave functions with and without Jahn-Teller 
distortion. In conclusion, there are triplet excitations which 
can move coherently throughout the lattice. 
 
The second coherent excitation we consider is charged. Let us
suppose we add (or remove) one electron, as in a 
photoemission experiment. First let us neglect electron hopping. The additional
electron modifies the occupancy of a single molecule leading to
a degenerate ground state with 
$l=0$, two electrons in the $d_2$ orbital, one
in the $d_1$ and $\lambda=\pm 1/2$. As for the exciton, also this charged 
excitation can move coherently, i.e. without changing the total energy
during the hopping process, and gives rise to two bands resulting
from the molecular degeneracy $\lambda=\pm 1/2$. The 
general Bloch wavefunction is
\[
\mid \vec{k}_{\pm} \rangle = \frac{1}{\sqrt{2 N}}
\sum_{\vec{R}} \left( \mid \vec{R},\lambda=+1/2\rangle\, 
\pm \mid \vec{R},\lambda=-1/2\rangle\right) {\rm e}^{-i\vec{k}\cdot\vec{R}} ,
\]
$\vec{R}$ being the position of the added electron and $N$ the number of sites.
The energy of this state is obtained by applying the hopping
Hamiltonian (\ref{hopp-d}) and is simply
\[
\epsilon_{\pm}(\vec{k}) = \frac{\Gamma}{2}
\sum_{\vec{r}} \left( t^{11}_{\vec{r}} \pm t^{12}_{\vec{r}}
\right) {\rm e}^{-i\vec{k}\cdot\vec{r}},
\]
where $t^{ab}_{ij} = t^{ab}_{i-j}$. We note that these bands
look exactly the same as those in the absence of electron-vibron coupling,
although with a bandwidth reduced by the factor $\Gamma$. This
is a clear manifestation of the dynamical JT effect. 
Moreover, it is important to notice that the bandwidth 
which results from such an ideal photoemission experiment
should not be identified with the
energy scale which compete against the intramolecular orbital energy
Eq.(\ref{onsite}) in deciding whether the Jahn-Teller distortion is static 
rather than dynamic. In fact, in the $U_{eff}\gg W$ limit we are
considering, that energy scale is instead $\propto W^2/U_{eff}$.
For smaller values of $U_{eff}$ (smaller, but still enough to make the system 
a Mott insulator), the value of this energy scale is expected to be 
perhaps larger than $W^2/U_{eff}$ but still smaller than $W$.
This point will be relevant to our discussion of section IV.      
Moreover, we stress that this result relies on a sort of 
adiabatic assumption, that is 
we have assumed that the molecule after the addition of the electron flows
into the corresponding ground state. Other configurations have been 
neglected since they are energetically disadvantageous even though
they might have larger matrix elements\cite{Manini&Doniach}. 

What is the spectrum of these charged excitations, instead, for a 
cooperative JT state? In the same adiabatic assumption we expect 
that the additional electron locally modifies
the Jahn-Teller deformation. It should then move coherently with this 
deformation, likely forming a single narrow polaron band instead of two
as in the dynamical JTM state.

Finally, a third class of coherent excitations involve only the
rotational degrees of freedom. In the limit $\omega_0\gg t^{ab}_{ij}$
each molecule is in a state with $\lambda=0$. The lowest excitations
consist of larger angular momentum eigenvalues. Let us excite a single
molecule into a state with $\lambda=1$. This excitation is highly
degenerate since $\lambda=-1$ has the same energy and also because
we can choose any of the molecules. These degeneracies are lifted by
the inclusion of the intermolecular exchange (\ref{Exc}). If we
restrict to the subspace where all molecules are in the 
$\lambda=0$ state but one, with $\lambda=\pm 1$, and diagonalize
(\ref{Exc}) in this subspace, we again find two dispersive bands for
this orbital excitation.
Orbital angular momentum is quenched in the cooperative Jahn-Teller
state. The presence of orbital excitations
would be strong evidence of the dynamical nature of the
Jahn-Teller effect, and of the JTM state.

\section{Are fullerenes A$_4$C$_{60}$ possible candidates?}

A possible candidate for the scenario we have presented could be,
at least in principle, the tetravalent 
fullerides $K_4 C_{60}$ and $Rb_4 C_{60}$. 
In fact, as we pointed out in the Introduction, 
this was the original source of inspiration of this work, even though, 
as we shell see, it could still be that the Jahn-Teller effect 
in these fullerides is in fact static and collective, rather than
dynamical. In these compounds the 
threefold-degenerate molecular orbital $t_{1u}$ of $C_{60}$ 
is partly occupied by the four electrons provided by the alkali atoms. 
In a purely band picture, and a rigid undistorted lattice,
this would imply a metal. In particular, the non cubic 
bct crystal field is not predicted to split sufficiently the degeneracy
so as to produce a band insulator\cite{bande}.
Yet these compounds are non magnetic insulators at ambient 
pressure\cite{resistivita,musr,ESR,optical-cond,spettroscopia} and
also undergo an insulator-to-metal transition under pressure\cite{Braz}.
A possible explanation of this behavior might just rely on the
mechanism we have till now discussed, that is a strong Coulomb repulsion
which drives the system into a Mott insulating phase, and a 
Jahn-Teller splitting which overcomes the Hund rule and 
makes the insulator non magnetic (see also Refs.\onlinecite{Braz,primadinoi}).
The estimated values of the
relevant physical parameters are not in contradiction with
this scenario. In fact, the bandwidth $W$ measured by photoemission is
of order 0.5-0.7 eV\cite{spettroscopia}, while the
Coulomb interaction $U$ may be a factor 1.5-2.5 larger than 
$W$\cite{U}. Moreover, single molecule calculations predict that the ground 
state of $C^{4-}_{60}$ is indeed a singlet due to prevailing 
Jahn-Teller effect, with an energy difference between the ground state
singlet and the lowest triplet of the order of 0.14 eV\cite{Manini}. 
Recent NMR measurements on K$_4$C$_{60}$\cite{Braz} show evidences of an 
exponential temperature dependence of the relaxation rate $T_1^{-1}$ 
with an activation energy comparable to the above mentioned splitting.    
Therefore, it seems that some of the properties of a single molecule
persist in the lattice, which is not incompatible with 
a Mott insulating state\cite{notaMott}.

\subsection{Isolated C$^{-n}_{60}$ molecules}

It is worthwhile to present a brief overview of some properties of 
the C$^{-n}_{60}$ molecular ions 
to point out the analogies with the simple model we have studied in the
preceding sections (we mostly follow Ref.\onlinecite{Manini}).  
Let us start by neglecting the intramolecular Coulomb repulsion.  
In the C$^{-n}_{60}$ molecule, the threefold 
degenerate $t_{1u}$ LUMO is Jahn-Teller coupled to eight fivefold 
degenerate $H_g$ vibrational modes.
We will discuss here the low energy picture at strong coupling ($g\gg 1$),
even though the realistic estimate is $g\sim 1$. As discussed in
Ref.\onlinecite{Manini}, the strong coupling analysis gives 
a qualitatively good description of the lowest energy excitations
also at $g\sim 1$, since no level crossing occurs above this value.
For strong coupling
the lowest energy modes correspond to those of a rigid body rotator
for $n=3$ and of a point particle moving on a sphere for $n=1,2,4,5$. 
Moreover, analogously to the model previously discussed,
the condition that the vibronic wave function be single valued
generates ``Berry's phase''\cite{Ihm,Manini} constraints on the quantum 
numbers of those lowest excitations. 
Specifically, for $n=3$ the rigid body rotator eigenfunctions are 
identified by the quantum numbers $L$, $L_z$ and $k$, where $L$ and
$L_z$ are the eigenvalues of the rotator top angular momentum 
$\vec{L}^2$ and its $z$-component, respectively,  and $k$ the 
eigenvalue of the rotation
around the corotating axis. These quantum numbers are subject to the
constraint $L=odd$ and $k=even$\cite{Manini}. This implies for instance that 
for $n=3$ the
molecular ground state is sixfold degenerate ($L=1$ and $S=1/2$). Here,
the JT effect does not remove completely the original orbital
degeneracy. On the other hand, for
$n=1,2,4,5$ the eigenfunctions are simply spherical harmonics
with quantum numbers $L$ and $L_z$ subject to the
constraint $(-1)^{L+n}=1$\cite{Manini}. Specifically for $n=4$,
which is the relevant case for Rb$_{4}$C$_{60}$ and K$_{4}$C$_{60}$,
the ground state is non degenerate ($L=0$ and $S=0$). The Jahn-Teller 
effect fully removes in this case the orbital degeneracy.
This JT energy gain is predicted to overcome the Hund's rule energy 
(favoring a high spin state) thus giving
a non degenerate ground state for $n=4$.
The lowest singlet excitations in the limit of strong JT 
effect correspond to the motion of a particle on a sphere.
They are therefore parametrized by solid angle coordinates $(\theta,\phi)$,
and are diagonalized in terms of spherical harmonics $Y_{LM}(\theta,\phi)$
with $L$ even. Their typical energy scale
is of order 0.02 eV\cite{frequenza}. The ground state configuration 
of each molecule is non degenerate with $L=M=0$, in spite of the degeneracy
in the absence of electron-vibron coupling.

\subsection{The lattice of C$^{-n}_{60}$ molecules}

As we said, the ratio of $U_{eff} = 0.5[E({\rm C}^{3-}_{60}) + 
E({\rm C}^{5-}_{60}) -2E({\rm C}^{4-}_{60})]$ and  
electron bandwidth $W$ is likely above the critical value for the
onset of the Mott insulating phase\cite{Gunn-Mott}.
In this case, the hopping is expected 
to produce just an intermolecular coupling which, as before, tends
to fix the angles $(\theta,\phi)$ of each molecule, thus contrasting the
intramolecular energy which favors the conjugate angular momentum
operator to acquire a finite eigenvalue. 
Moreover, one should take into account additional effects which tend to fix
the phases, like the crystal field and multipolar forces among the molecules. 
The interplay of all these terms
to the total Hamiltonian might give rise in this case too to the two
different kinds of insulating states just discussed, i.e.
the collective static Jahn-Teller and the dynamical Jahn-Teller Mott
insulators.
However, the order of magnitude of the various relevant parameters
seem at present to point more in favor of a static and collective distortion. 
In fact, 
the intermolecule coupling arising from the electron hopping is expected to 
be of the order $W^2/U_{eff}\sim 0.2 eV$, that is an order of magnitude larger
than the rotational lowest excitation energy. In addition, one has to consider
the crystal field, which also favors a static distortion.
In the end, in the tetravalent fullerides, this issue will have to be resolved
experimentally. In fact, at least to our knowledge, there is so far  
no clear experimental 
evidences in favor either of 
a cooperative and static, or of a dynamical Jahn-Teller 
Mott-like insulator. Statically, the bct
lattice being bipartite, we might expect a staggered molecular distortion
doubling the lattice unit cell 
to be more favorable than a homogeneous one. 
In fact, S. Erwin has predicted on the basis of a band structure LDA 
calculation a CDW instability with wave vector $\vec{Q}=(0,0,\pi/c)$,
due to a Fermi surface quasi-perfect nesting\cite{bande}. This would lead to 
a doubling
of the bct unit cell along the c-axis, for which there is as yet no evidence
in X-ray diffraction experiments\cite{Xray}. In the Jahn-Teller-Mott state,
no such doubling is required.

\section{Conclusion}
In this paper we have investigated the properties of a prototypical
Jahn-Teller insulator, consisting of a lattice of molecules.
Each molecule has a degenerate electronic orbital (the degeneracy being 
bigger than half the number of valence electrons, which is taken as even), 
coupled to a degenerate vibron, and can undergo a Jahn-Teller distortion.
For the case of degeneracy two, which we treated explicitly,
the molecular ground state with two electrons 
is non degenerate due to the Jahn-Teller effect overcoming
the Hund's rule splitting, which would favor a high-spin state instead. 
If a lattice of
such doubly occupied molecules is constructed, and if the effective
Hubbard repulsion overcomes the electron hopping matrix elements, the system
is insulating. However the insulating state has different properties
depending whether a cooperative and static Jahn-Teller distortion 
is realized or whether instead a dynamical, non-magnetic Jahn-Teller-Mott
state prevails. By analyzing the
low lying excitations, we have identified some characteristic 
features which could discriminate between the two insulators.
Specifically we have shown that these excitations have some reminiscences
of the original degeneracy of the electronic
as well as of the vibronic degrees of freedom in the dynamical case 
which are lost if a cooperative Jahn-Teller effect takes place.  

We have also briefly discussed the possible occurrence of this Jahn-Teller Mott
insulator in the fullerides K$_{4}$C$_{60}$ and Rb$_{4}$C$_{60}$. 
Apart from the 
obvious differences arising from the larger degeneracy of the
electronic orbitals and of the vibronic modes involved, 
we believe that the qualitative picture obtained in the simpler
doubly degenerate model is useful to understand this case. 
Although the order of magnitude of the relevant physical parameters
seem to favor a scenario in which a static and cooperative
Jahn-Teller distortion takes place,  
so far there are to our knowledge no experimental data which clearly point in
favor of this situation rather than a dynamic Jahn-Teller effect. 
Therefore the question for these systems remains open and, 
to our opinion, is interesting enough to deserve further experimental 
investigations.

\section{Acknowledgments}
This work has been partly supported by the EEC, under contract 
ERB CHR XCT 940438, and by INFM, project HTSC.

\begin{table}

\begin{tabular}{lcr}
n & quantum numbers & $E(r)$ \\ \tableline
1 & $s=1/2$, $\tau_x=-1/2$ & $U + V/2 - \omega_0 g r/2 + 
\omega_0/(8r^2)$ \\ 
1      & $s=1/2$, $\tau_x=1/2$  & $U + V/2 + \omega_0 g r/2 + 
\omega_0/(8r^2)$ \\ 
2 & $s=0$, $\tau_x^2=1$ & $4U + 3V/2 + \omega_0/(4r^2)
- \sqrt{V^2/4 + \omega_0^2 g^2 r^2}$ \\ 
2      & $s=1$, $\tau=0$     & $4U$ \\ 
2      & $s=0$, $\tau_x^2=0$ & $4U + V + \omega_0/(2r^2)$ \\ 
2      & $s=0$, $\tau_x^2=1$ & $4U + 3V/2 + \omega_0/(4r^2)
+ \sqrt{V^2/4 + \omega_0^2 g^2 r^2}$ \\ 
3 & $s=1/2$, $\tau_x=-1/2$ & $9U + V/2 - \omega_0 g r/2 + 
\omega_0/(8r^2)$ \\ 
3      & $s=1/2$, $\tau_x=1/2$  & $9U + V/2 + \omega_0 g r/2 + 
\omega_0/(8r^2)$ \\
\end{tabular}
\caption{Electronic eigenstates and eigenvalues at fixed $r$,
for occupation numbers $n=1,2,3$}
\end{table}

\newpage 
\begin{figure}
\centerline{\epsfbox{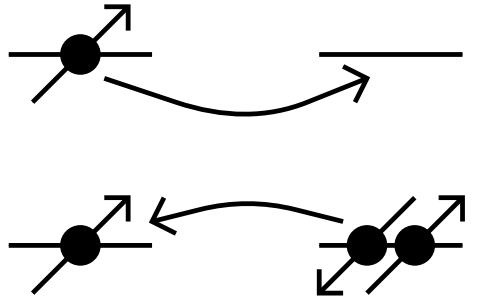}}
\caption{Second order process responsible of the motion of the $S=1$ 
excitation}
\end{figure}


\begin{references}
\bibitem{notax}The $\tau_z$ dependent term in Eq.(\ref{Hambis})
can be approximated at order $g^{-2}$ by
\[
\left( \frac{\partial}{\partial \phi} + i\tau_z\right)^2
\simeq  \frac{\partial^2}{\partial \phi^2} - \frac{1}{2}(
\vec{\tau}\cdot\vec{\tau} - \tau_x^2).
\]
\bibitem{notaMott}Notice that a simple uniform Jahn-Teller distortion
can not explain the insulating behavior, since the bandwidth is 
larger than the Jahn-Teller contribution to $U_{eff}$ (the latter being
estimated of the order 0.2-0.4 eV). Therefore the Coulomb repulsion
has to play a relevant role in making the system insulating.
\end{references}
\end{document}